# Self-selective growth of $GaAs_{1-x}Bi_x$ on GaAs zinc blende/wurtzite nanowire heterostructures

Rohit Yadav[1,2,¤,*], Sebastian Lehmann[2,3], Vidar Flodgren[1,2], Evangelos Golias[4], Alexei Zakharov[4,†], Kimberly A. Dick[2,3], Anders Mikkelsen[1,2], and Rainer Timm[1,2,#]

[1]Division of Synchrotron Radiation Research, Lund University, 221 00 Lund, Sweden
[2]NanoLund, Lund University, 221 00 Lund, Sweden
[3]Centre for Analysis and Synthesis, Lund University, 221 00 Lund, Sweden
[4]MAX IV Laboratory, Lund University, 221 00 Lund, Sweden
[¤]Present Address: NNF Quantum Computing Programme, Niels Bohr Institute, University of Copenhagen, 2100 Copenhagen, Denmark

*E-mail: rohit.yadav@nbi.ku.dk
[†]deceased
[#]E-mail: rainer.timm@fysik.lu.se

## Abstract

Site-selective nanostructure growth and material incorporation at the atomic scale offer a promising pathway for engineering quantum materials and nanodevices. Here, GaAs nanowires (NWs) with an axial heterostructure of alternating zinc blende (Zb) and wurtzite (Wz) crystal phases are employed as templates for site-selective Ga and Bi overgrowth. Using X-ray photoemission electron microscopy (XPEEM) with nanoscale spatial resolution, we map elemental distribution and local chemical bonding to reveal the incorporation behavior of Bi atoms in {110} Zb and $\{11\bar{2}0\}$ Wz facets. Bi incorporation proceeds through an anion-exchange process, where Bi atoms replace As, forming local Ga-Bi bonds and producing a thin $GaAs_{1-x}Bi_x$ shell. We observe crystal-phase-dependent Bi incorporation, with higher Bi concentration in the Zb segments than in the neighboring Wz segments within the same NW. Furthermore, the Zb segment with higher Bi content exhibits reduced susceptibility to oxidation compared with the Wz segment, resulting in increased Ga-oxide in the Wz surfaces. This study highlights GaAs NW Zb/Wz heterostructures as a template for controlled growth of GaBi and $GaAs_{1-x}Bi_x$ nanostructures with tailored functionalities for quantum applications.

## Introduction

Controlled and site-selective growth of nanostructures is a fascinating challenge, offering promising routes towards functional nanodevices and providing a platform for studying quantum phenomena. It can be achieved by post-growth atom manipulation or by nucleating nanostructures on a suitable, typically pre-patterned substrate. The former has been applied to studying single-atom transistors,[1] quantum dots,[2] and quantum rings[3] on semiconducting surfaces. Although these approaches are fascinating for fundamental science, they are not suitable for large-scale industrial processing. More scalable post-growth processes implementing steps like lithography or etching have been reported, such as for nanoscale InAs- and graphene-based transistors.[4,5] However, they are limited by the spatial resolution of fabrication techniques, such as electron/optical lithography. The latter approach of nucleation on a suitable substrate can be a promising alternative to the post-growth approaches. It is scalable, limited by the deposition rate of the adsorbate or by the substrate pre-treatment conditions. Some of the important examples include atomic chains of metal on semiconducting planar surfaces,[6] including self-assembled atomic chains of Bi on InAs(110),[7] which, however, lack spatial control of the chain formation, and approaches based on substrate strain engineering[8] or cleaved-edge overgrowth[9] requiring complex processing steps.

A very elegant alternative is to use III-V semiconductor nanowires (NW) with axial crystal phase heterostructures as a template for site-selective overgrowth. Such nanowires, with diameters between about 50 and 500 nm and lengths of up to several microns, are unique because they allow for controllable switching between cubic zinc blende (Zb) and hexagonal wurtzite (Wz) crystal phases upon axial NW growth,[10] while the Wz phase is metastable during bulk growth of most III-V compound semiconductors. The atomic structure and various surface facets of these Zb/Wz NW heterostructures have been well studied using atomic-scale techniques like scanning tunneling microscopy/spectroscopy (STM/S).[11,12] We have previously used GaAs Zb/Wz NW heterostructures to incorporate Bi or Sb atoms into the surface facets and observed different incorporation probabilities for Zb and Wz facets due to the different atomic surface structures.[13,14] This was supported by density functional theory (DFT) calculations of diffusion barriers and the energy balance of Sb-for-As exchange processes at the different facets.[13] While those Bi deposition studies were performed in the well-controlled environment of an ultrahigh vacuum (UHV) STM chamber, allowing for immediate atomic-scale characterization, a more scalable epitaxy approach would be needed for larger-scale application of $GaAs_{1-x}Bi_x$ heterostructures.

Probing Bi incorporation is particularly important due to its large spin-orbit splitting and reported low-dimensional quantum states.[15] Furthermore, III-bismide species like GaBi and InBi are expected to show non-trivial topological insulator behavior and induce band inversion, which is large enough to enable room-temperature applications.[16,17] Bulk growth of binary III-Bi compounds has not been realized yet, and even the growth of III-Bi structures on planar III-V semiconductor surfaces using conventional epitaxial techniques is challenging. For example, Bi deposition on polar III-V surfaces like InSb(111), InAs(111), or GaAs(111) typically results in Bi-terminated surface reconstruction, [18–21] while on GaAs(110), it has been reported to result in island growth.[22,23] However, when deposited on the Wz {11-20}-type surfaces present on GaAs NW, Bi was found to form ordered 1D GaBi chains and 2D GaBi nano islands, as studied using STM measurements.[14] Although a thorough investigation of the chemical interaction of Bi with the GaAs surface of the NW underneath is still lacking. On planar surfaces, such investigations are typically performed using X-ray photoelectron spectroscopy (XPS).[18–21] However, its application to individual NWs is challenging because the X-ray footprint in conventional XPS is typically on the micrometer scale. To characterize the chemical composition of the NW surfaces, we employed X-ray photoemission electron microscopy (XPEEM), which provides element-specific real-space imaging with spatial resolution down to 20 nm.[24–26] Investigating stable epitaxial growth of GaBi or $GaAs_{1-x}Bi_x$ films on GaAs NW surfaces beyond the initial Bi-induced nucleation stage and correlating NW structure, overgrowth, and chemical composition at the nanoscale has not yet been achieved and is a significant step toward engineering functional semiconductor materials.

Here, we have studied GaAs Zb/Wz NW heterostructures radially overgrown with a thin layer of Ga and Bi. Using XPEEM, we examine the incorporation mechanism and chemical characteristics of Bi atoms into the {110} Zb and $\{11\bar{2}0\}$ Wz facets of the GaAs NW. We reveal distinct Bi incorporation behavior between Zb and Wz segments, with enhanced Bi incorporation in Zb regions compared to neighboring Wz segments. The incorporation process is found to be explained by the Bi-for-As anion exchange reaction, resulting in prominent Ga-Bi bonding and formation of a thin $GaAs_{1-x}Bi_x$ shell. In addition, the Wz segments exhibit stronger oxidation than the Zb ones, suggesting that Bi incorporation reduces surface oxidation. These findings provide insights into crystal-phase-selective overgrowth of GaAs Zb/Wz NWs, and they provide a path towards controlled growth of GaAs/GaBi nm-scale heterostructures.

## Method

**GaAs NW growth:** The design of the Ga-As-Bi nanowire samples was divided into two growth steps, one for the Wz–Zb heterostructured GaAs templates and one where Ga and Bi were simultaneously deposited onto these pregrown templates from step one. Following the vapor-liquid-solid growth mode[27] the template GaAs NWs were grown by metal organic vapor phase epitaxy (MOVPE). Au aerosol particles with a nominal diameter of 50 nm and an areal density of 34 $\mu m^{-2}$ were deposited onto GaAs $(\bar{1}\bar{1}\bar{1})$ substrates using a technique outlined previously.[28] These prepared growth substrates were loaded into a 3×2” Aixtron close-coupled showerhead reactor (CCS) where MOVPE growth of the GaAs NWs was carried out at a total reactor pressure of 100 mbar, using hydrogen as carrier gas, and at a total reactor flow of 8 slm. Before initiating the actual growth, the substrates were annealed at a set temperature of 630 °C for 10 min in an arsine ($AsH_3$)/ hydrogen ($H_2$) atmosphere, to remove surface oxides and allow for proper surface conditioning. After lowering and stabilizing the set temperature at 550 °C, nanowire growth was initiated by introducing trimethylgallium (TMGa) and $AsH_3$ with molar fractions of $\chi_{TMGa} = 1.9\times10^{-5}$ as well as $\chi_{AsH3} = 4.5\times10^{-5}$ and $4.4\times10^{-3}$ for Wz and Zb conditions, respectively. Heterostructured NWs consisting of alternating four Zb segments embedding three Wz segments with lengths of the Zb and Wz segments in a range from approximately 10 to 200 nm were engineered by only switching the group V precursor flow. For detailed information about the growth of sharp crystal structure interfaces in III–V NW systems, the reader is kindly referred to Ref.[29] and references therein. The growth step was terminated by cutting the TMGa flow and cooling in an $AsH_3/H_2$ atmosphere down to 300 °C before also cutting the $AsH_3$ flow. The so grown substrates with nanowire templates were treated for 2 minutes in an ultrasound bath in a water/isopropanol (30% vol.) solution to only leave NWs with diameters larger than approximately 200 – 400 nm on the substrates. After that step the samples were reintroduced into the growth chamber for the second growth step. After identical settings and times for the conditioning and oxide removal, the set temperature was lowered to 400 °C where, after a temperature stabilization step, TMGa and Trimethylbismuth (TMBi) were supplied for 2 minutes with molar fractions of $\chi_{TMGa} = 3.9\times10^{-5}$ and $\chi_{TMBi} = 1.1\times10^{-5}$. Closing both supplies simultaneously and cooling down under $H_2$ only terminated the growth step. The GaBi layer is then capped with a GaAs layer, using the precursors and procedures mentioned above. The schematics of the grown NW and scanning electron microscopy (SEM) images are shown in Fig. 1. The SEM images were recorded in a ZEISS Gemini 1560 microscope equipped with a field emission gun and operated at 15 kV. Underneath and above the Zb/Wz heterostructure, a long NW growth stem of pure Wz phase and a short top segment, terminated with the catalytic Au/Ga alloy particle, can be seen for most NWs.

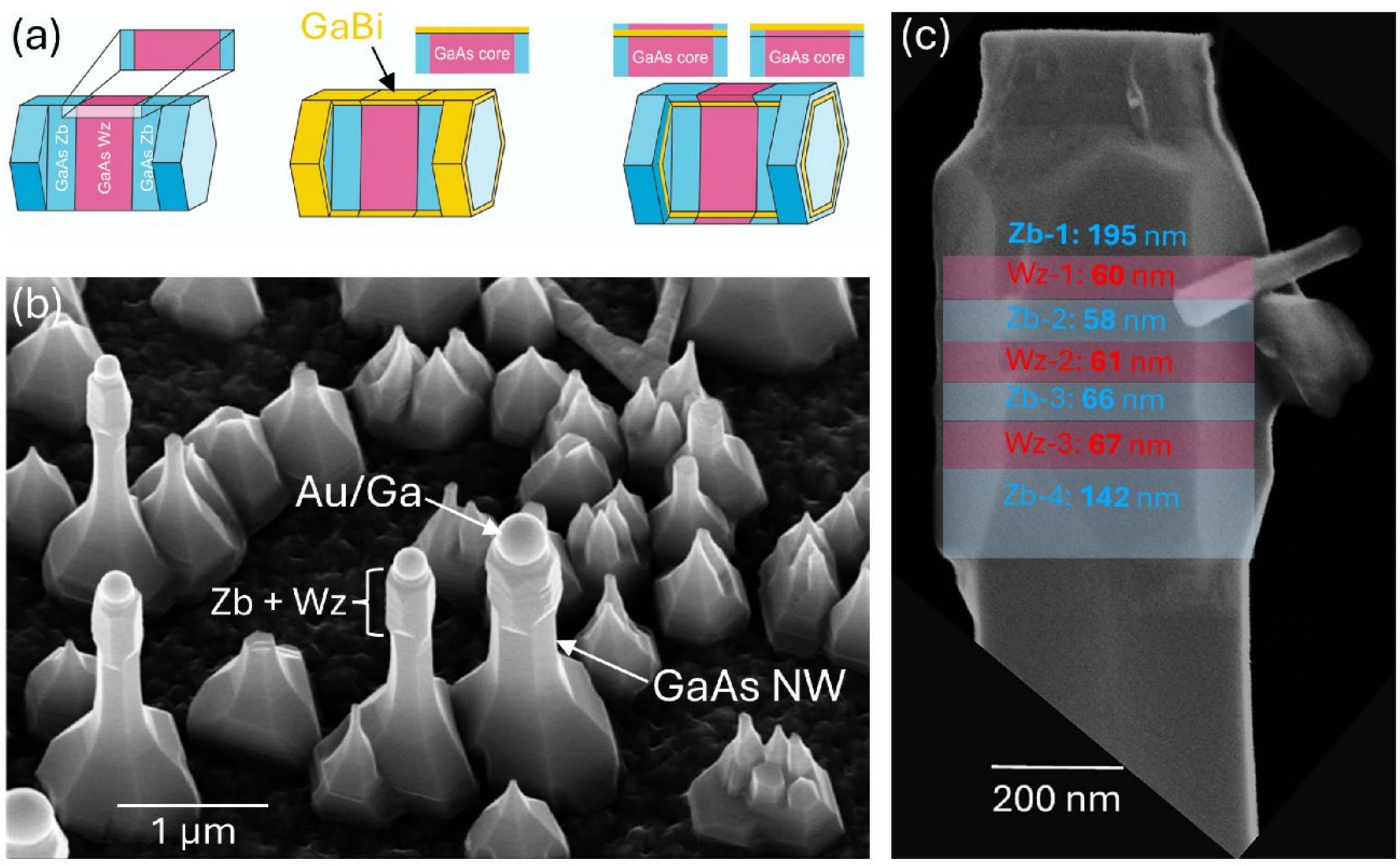

***Figure 1*** *(a) Schematic representation of the GaBi/GaAs NW growth with Zb and Wz heterostructure; SEM image of (b) grown NWs on the growth substrate; (c) a transferred NW on the patterned Si substrate for XPEEM measurement, color coded for identifying heterostructures, red for Wz and blue for Zb.*

**NW transfer:** The heterostructure NWs were transferred using a small piece of cleanroom-compatible paper under ambient conditions. A small area of the growth sample was dabbed, and the picked-up NWs were transferred onto dedicated finder chips with Ti/Au reference markers on silicon. The reference markers were used as a coordinate system for sample alignment on the substrate. For this, we used the open-source integrated-circuit (IC) layout editor software KLayout and its versatile Python API to automatically construct a purpose-made UV lithography (UVL) mask. The design consists of a tileable 2×2 cm unit cell, each consisting of a row-column labelled coordinate grid with center-cross marker separations of 50 microns. When tiled across a 2” Si wafer, each chip is also given a title label that allows for quick identification in a microscope.

Before processing, we must clean the Si wafer. This was done by sonication, once in acetone, and twice in isopropyl alcohol for 60 s each, after which it was prebaked at 165 °C for 5 minutes. For UVL, we use a conventional resist bilayer consisting of LOR10b and S1813. LOR10b was spin-coated at 3000 rpm for 60 s, followed by baking at 165 °C for 5 minutes. S1813 was spin-coated at 3000 rpm for 60 s, followed by baking at 115 °C for 90 s. UVL mask exposure was done using a Heidelberg MLA 150 maskless aligner and its 405 nm laser, with a dose of 150 $mJcm^{-2}$ and a +0 defocus. Following exposure, the wafer was developed for 50 s in MF319. A quick inspection under an optical microscope confirmed if development needed more time, and if not, then $O_2$ plasma ashing was done for 15 s to remove biological residues. Following this, contact metallization was done using a Temescal E-beam evaporator (20/200 nm Ti/Au), with lift-off being done by the submersion of the wafer in acetone until completion. Finally, the wafer was diced into 2×2 cm device chips using a Disco DAD 3320.

**Sample cleaning and annealing:** The samples with transferred NWs were mounted on dedicated XPEEM sample holders and exposed to a flow of clean nitrogen gas to remove dust or particles. After transferring into the UHV chamber of the XPEEM setup, the samples were annealed in the presence of atomic hydrogen for native oxide removal, following previous experience.[12,14] Atomic hydrogen was provided using a commercial thermal cracker (from MBE Komponenten) with a molecular hydrogen pressure of 2 x $10^{-6}$ mbar at a cracking temperature of 1700 °C.

**X-ray PEEM measurement:** The XPEEM measurements were carried out at the MAXPEEM beamline of the MAX IV Laboratory (Lund, Sweden) using an Elmitec aberration-corrected Spectroscopic Photoemission and Low Energy Electron Microscope (SPELEEM). Unless otherwise specified, a photon energy of 80 eV was used for core-level maps. Before analyzing the XPEEM results, the images were drift-corrected using the Igor Pro analysis package Athina.[30]

## Results

The NW samples were annealed at 260 °C in the presence of atomic hydrogen for 10 minutes, to remove some of the native oxide from their surfaces (without risking desorbing Bi, which would occur at too high annealing temperatures).[31] After that step, the NWs were excited by synchrotron X-rays, and XPEEM images were collected to obtain core-level maps. Figure 2(a,c,e) shows XPEEM images obtained at a binding energy (BE) of 24.2 eV, 20.0 eV, and 41.6 eV, which were chosen to extract Bi $5d_{5/2}$, Ga *3d*, and As *3d* core-level maps, respectively. In all three maps, a striped pattern can be seen, where the same three NW segments appear dark in the Bi map and bright in the Ga and As maps. By comparing the XPEEM results with an SEM image of the same NW (imaged after XPEEM characterization) and the nominal growth structure (cf. Fig. 1(c)), it becomes evident that the three stripes of the same contrast correspond to the three Wz segments of the NW heterostructure, surrounded by Zb segments. In addition, one can see a long, straight NW segment underneath the heterostructure, which is the NW growth stem, and a shorter NW segment above the heterostructure, while the Au/Ga alloy particle seems to have been misplaced from the top of the NW during transfer. Intensity profiles are taken from the Bi and Ga XPEEM maps, across the NW heterostructure, shown in Fig. S1 of the electronic Supplementary Information (ESI). The extension of the Wz segments, appearing bright in the Ga *3d* map, and of the Zb segments in between, appearing bright in the Bi *5d* map, both amounting to 40 to 60 nm, is consistent with the corresponding values obtained from the SEM image, shown in Fig. S1.[26]

From the XPEEM maps, the observed different contrast for the two different crystal phases of the NW indicates different overgrowth and Bi incorporation mechanisms. A higher Bi *5d* intensity is observed from the Zb phase compared to the Wz, as shown in Fig. 2(a), although it is not zero in the Wz phase. The opposite trend is observed in the Ga *3d* XPEEM map, where a higher intensity is found from the Wz phase compared to Zb (Fig. 2(c)). For a more quantitative analysis, a series of additional XPEEM maps were recorded with varying BE, centered around the BE values of the maps shown in Fig. 2. A resulting map of binding energy vs distance along the NW (across the axial heterostructure) is plotted in Fig. S2 of the ESI. From the series maps, Bi *5d,* Ga *3d*, and As *3d* core level XPS spectra can be extracted at the center of Wz and Zb segments. The corresponding photoemission spectra are plotted in Fig. 2(b,d,f). Note that the energy step width, and thus the effective energy resolution, depends on the number of XPEEM maps acquired in the energy series. It is therefore not directly comparable to the energy resolution of XPS data acquired at dedicated beamlines, which, however, lack the nanoscale spatial resolution provided by XPEEM. As can be seen in Fig. 2(b), the intensity of the Bi $5d_{5/2}$ peak at the Zb segments is significantly higher than that at the Wz segments, suggesting higher Bi incorporation into the Zb phase material as compared to the Wz phase.

Previously, we have deposited sub-monolayer amounts of atomic Bi onto the surfaces of GaAs NW Wz/Zb heterostructures and reported the incorporation of Bi atoms into the GaAs surface, replacing As in the group-V lattice sites, explained by a Bi-for-As exchange reaction.[14] In that study, the Zb {110} surfaces generally showed a higher density of Bi atoms than the Wz $\{11\bar{2}0\}$ surfaces, even though the exact ratio depended significantly on NW geometry and surface step density. A similar behavior was also observed in the case of Sb deposition on GaAs NWs, where the higher probability of Sb incorporation in Zb GaAs NW surfaces as compared to Wz GaAs NW surfaces was supported by DFT results, suggesting that the group-V exchange reaction (in that case Sb-for-As exchange) was energetically more favorable on Zb {110} GaAs as compared with Wz $\{11\bar{2}0\}$ GaAs.[13] Interestingly, a similar behavior is also observed here, even though in this work, Ga and Bi atoms were offered simultaneously during MOCVD growth. The anion exchange reactions have previously also been observed in similar planar systems.[14,31,32] Thus, a Bi-for-As exchange can be expected not only upon deposition of atomic Bi on the GaAs surface, but also in our case of NW shell growth by MOCVD.

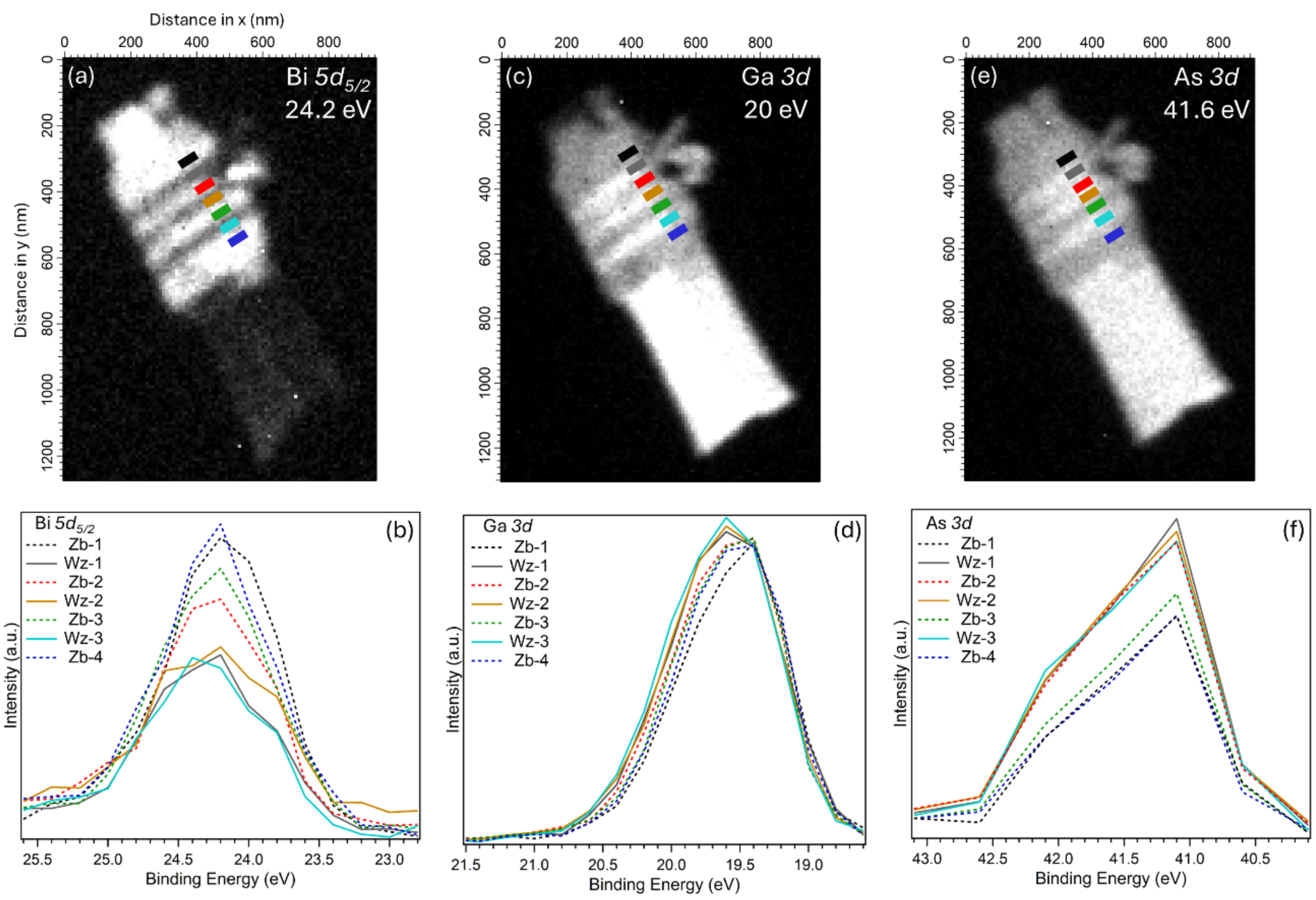

***Figure 2*** *XPEEM (a) Bi 5d, (c) Ga 3d, and (e) As 3d core-level maps of the same NW, and corresponding (b) Bi* $5d_{5/2}$*, (d) Ga 3d, and (f) As 3d core-level spectra, extracted from series of XPEEM maps with slightly varying BE, for different color coded regions in (a,c,e) corresponding to the Wz and Zb segments of the NW. The photon energy used for XPEEM mapping is 80 eV. The NW growth structure is illustrated in Fig. 1.*

The Ga *3d* XPEEM map presented in Fig. 2(c) shows an opposite contrast pattern of the Wz/Zb heterostructure compared to the Bi *5d* map in Fig. 2(a). A more complex behavior can be found in the Ga *3d* core-level spectra, which are compared in Fig. 2(d), obtained from different Wz and Zb segments. Interestingly, here the peak intensity of the Ga spectra is similar throughout the heterostructures, but the BE peak position and the shape of the spectra vary; spectra obtained at the Wz segments are broader and extend further to higher BE as compared to the Zb spectra, with the peak position shifted by 0.2 to 0.3 eV. Ga-oxide components are reported to have higher BE than the Ga-As component in Ga *3d* spectra.[33] Thus, we assign the broader peak from Wz to be due to the presence of a Ga-oxide component (along with other components like Ga-As and Ga-Bi), while the Zb spectra contain no or smaller Ga-oxide components. Oxidation of the NWs and changes in the XPS data upon oxide removal will be further discussed later. A slight asymmetry of all peaks with an increased intensity shoulder towards higher BE is due to the doublet nature of the Ga *3d* peak with a spin-orbit splitting of 0.45 eV. We conclude that the Wz and Zb segments contain a similar amount of Ga at the surface, only slightly higher in the Wz segments, according to the integrated Ga *3d* XPS intensity, except the Wz segments are more oxidized. It should be noted that the Ga *3d X*PEEM map shown in Fig. 2(c) is obtained at 20.0 eV BE, which is at the high BE side of the Ga *3d* spectrum, resulting in an increased sensitivity for the Ga-oxide component in the map. An XPEEM map for lower Ga *3d* BE (19.4 eV), which is mainly sensitive to the Ga-As and Ga-Bi components, presented in Fig. S2(a), indeed shows almost identical XPS intensity for the Wz and Zb segments of the NW. In addition, a map of binding energy vs distance along the NW for the Ga 3d core level, obtained from a series of Ga *3d* XPEEM maps with varying BE, is shown in Fig. S2(c).

The As *3d* XPEEM map and the corresponding spectra from Wz and Zb segments are shown in Fig. 2(e,f). The As *3d* spectra show no indication of an oxide component, which again would be in the high BE region,[33] therefore, we assume that most of the native As-oxide was removed from the NW upon annealing. However, like Bi *5d,* the As *3d* signal also shows an intensity variation between the Wz and Zb crystal phases. In contrast to Bi, the As *3d* intensity is higher in most regions of the Wz phase compared to the Zb phase.

Based on the above arguments and simultaneously comparing Bi *5d*, Ga *3d*, and As *3d* core-levels data, the following mechanisms for Bi incorporation into the GaAs NW surface and NW shell growth can be proposed: Bi atoms preferentially incorporate into the NW Zb segments via anion exchange reaction, replacing the As atoms and bonding to Ga. This results in both the observed lower As content and higher Bi content in the Zb segments, as compared with Wz. We want to point out that during MOCVD NW shell growth, not only Bi, but also Ga was provided, which nominally should lead to GaBi growth and not only Bi-for-As exchange at the NW surface. However, by comparing the intensities of Ga *3d* and As *3d* core-levels, and taking into account the photoionization cross section, which at a photon energy of 80 eV is about two times larger for Ga *3d* as compared to As *3d,*[34] we can exclude any significant attenuation of the As signal due to a thick, epitaxially grown GaBi layer. While we cannot fully exclude direct epitaxial GaBi growth, we can estimate the upper limit of the resulting film thickness to a few Å. This is in line with the known challenges in III-Bi bulk growth and also with the surfactant nature of Bi.[35] The different probability of the anion exchange reaction on Wz and Zb GaAs[13] is considered as the origin of the larger Bi and smaller As content on the surfaces of the Zb segments. In other words, the MOCVD overgrowth with both Ga and Bi leads to the formation of a thin $GaAs_{1-x}Bi_x$ shell around the GaAs NWs, with the Bi content being larger for Zb and smaller for Wz segments.

Most of the incorporated Bi atoms are supposed to sit at or close to the NW surface, where they influence the oxidation behavior. The studied NWs have oxidized after MOCVD growth upon transfer through air, and most of the native surface oxide has been removed again upon annealing in the UHV chamber of the XPEEM setup. It has been observed before that As-oxide components are easier to remove than Ga-oxide[12] in agreement with the results of the As *3d* and Ga *3d* spectra obtained here. As mentioned above, the Ga *3d* spectra show a larger amount of Ga-oxide on the Wz surface facets as compared to the Zb surface (cf. Fig. 2(d)), which implies that either Ga-oxide is more difficult to remove from the Wz surface upon annealing, or that less Ga-oxide forms on the Zb surface upon exposure to air. Importantly, the density of Bi atoms at the NW surface is higher for the Zb segments and lower for Wz. While this difference is expected to be less relevant upon the removal of Ga-oxide, it seems to have a significant impact on surface oxidation. We interpret the results by GaBi being more stable against the formation of Ga-oxide upon exposure to ambient atmosphere than GaAs, resulting in a smaller amount of surface Ga-oxide on those NW segments that have a higher Bi content. We tested this hypothesis by further annealing the NWs at a higher temperature of 300 °C for 10 min in the presence of atomic hydrogen. Figure 3(a) shows X-ray photoelectron spectra generated over a larger BE energy range, including both the Bi $5d_{5/2}$ and the Ga *3d* core-levels, obtained from the same NW as before (cf. Fig. 3), but after the extra annealing step. The variation in the Bi signal intensity is evident and the same as before the annealing (cf. Fig. 2(b)), while the Ga spectra look almost the same for most segments of the heterostructure, showing no difference anymore between Zb and Wz segments. Instead, all Ga *3d* spectra now show the same narrower peak that previously was only observed at the Zb segments. Accordingly, most of the previously found Ga-oxide has been removed upon annealing at higher temperatures. Considering the increased amount of Ga-oxide at the Wz segments after low-temperature annealing, this behavior confirms that the Wz segments, containing fewer Bi atoms, become more oxidized upon exposure to air and require more severe annealing to remove the oxides again. On the other hand, the difference in Bi content between the Zb and Wz segments remains even after the extra annealing step. We cannot exclude that some of the Bi atoms were desorbed upon annealing at higher temperatures, but the relative Bi intensity variation between Zb and Wz segments remained unchanged, demonstrating that this variation indeed comes from different amounts of Bi incorporated in the different segments and is not due to any oxidation behavior.

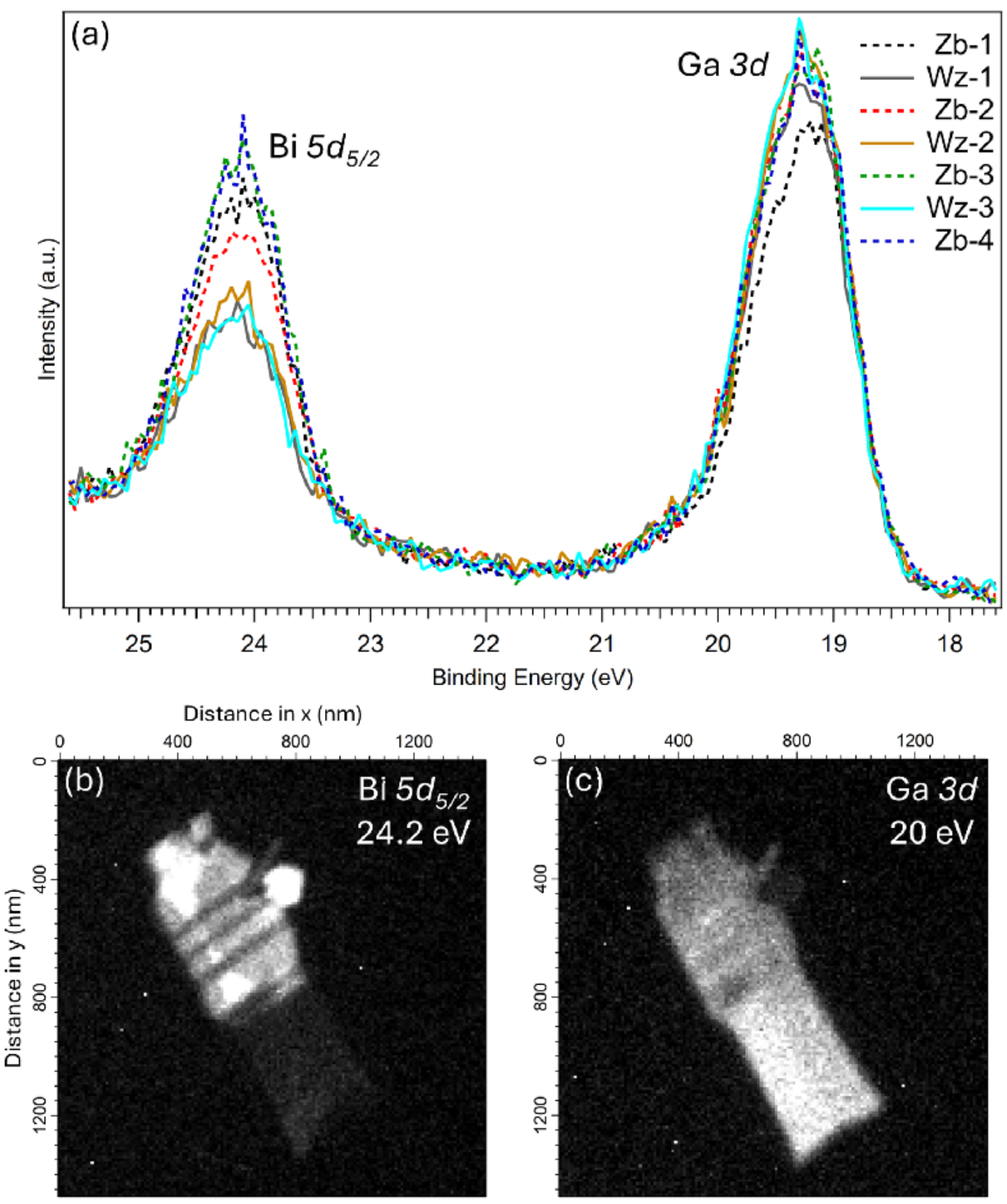

***Figure 3*** *(a) Combined Ga 3d and Bi 5d core-level spectra obtained at the same positions as those shown in Fig. 2(a,c), extracted from a series of XPEEM maps, including those shown in (b,c), covering a larger energy range. (b) Bi 5d and (c) Ga 3d XPEEM maps. XPEEM data were obtained after annealing at a higher temperature (300 °C), at 80 eV photon energy.*

The chemical composition of the NW surface and the type of bonds between Ga, Bi, and As atoms were further investigated by fitting Ga *3d* core-level spectra obtained from Wz and Zb segments, as shown in Fig. 4. The spectra are part of those shown in Fig. 2(d), and they were fitted with three Ga components. Fitting parameters and resulting peak BEs and relative intensities are shown in Tables S1 and S2 of the ESI, respectively. Even though the accuracy of the fits is limited due to the low energy resolution, we obtain qualitatively meaningful and robust results. We have assigned two main components to Ga-As (green), corresponding to the NW core material, and Ga-Bi (yellow), after Bi incorporation through Bi-for-As exchange. The BE difference between the Ga-As and Ga-Bi peaks is ca. 0.28 eV, which agrees well with previously reported results for Bi-induced surface structures on polar GaAs surfaces.[36] The fitted Ga-Bi component is larger for the Zb segment and smaller for the Wz segment (cf. Table S2), in agreement with the larger amount of Bi observed in the Zb surface. An additional component at higher BE is needed to fit the spectra completely; it is assigned to Ga-oxide (blue), in agreement with the literature.[37] The Ga-oxide component is significantly smaller on the Zb segment than on the Wz phase, as explained above. The relatively small chemical shift of 0.3 to 0.4 eV between the Ga-As and the Ga-oxide components indicates a $Ga^{+1}$ oxidation state for the Ga-oxide, as in $Ga_2O$. Here, one has to remember that the NWs have already been annealed, which may have reduced $Ga^{+3}$ oxides as in $Ga_2O_3$, which are easier to remove than $Ga^{+1}$ oxides.[38]

To further understand the chemical bonding configuration, we have also analyzed the Bi $5d_{5/2}$ spin-orbit coupling component, although peak fitting is difficult here due to strongly overlapping peaks and the low energy resolution. The large peak width of the Bi $5d_{5/2}$ singlet, which can be seen in the spectrum of Fig. 2(b) and Fig. 3(a), indicates that it is a convolution of at least 2 components. Considering that our NW shell consists of $GaAs_{1-x}Bi_x$ (with higher Bi content in the Zb segments and lower in Wz), we expect that the local Bi density varies at the atomic scale, and suggest that accordingly the BE of the Bi-Ga component varies slightly, depending on the distance to neighboring Bi atoms and therewith the local charge environment. Alternatively, one might argue that the broader Bi *5d* peak was due to a combination of Bi-Ga and Bi-Bi bonds, even though these two components have been reported to be indistinguishable[36] To evaluate this possibility of the existence of a significant amount of metallic Bi at the sample surface, we compare the relative intensities of the Bi *5d* and Ga *3d* spectra shown in Fig. 3a. Under the hypothetical assumption of a surface Bi layer that attenuates the signal from the GaAs substrate below, we use the XPS overlayer model,[39] as described in the ESI, with parameters given in Table S3, and obtain a calculated Bi layer thickness on the Zb and Wz segments of only ca. 1.9 Å and 1.5 Å, respectively (the used parameters are shown in Table S3). Thus, the possibility of a significant metallic Bi surface layer can be neglected, further supporting our result of Bi being incorporated through anion exchange, forming Ga-Bi bonds and resulting in as $GaAs_{1-x}Bi_x$ NW shell.

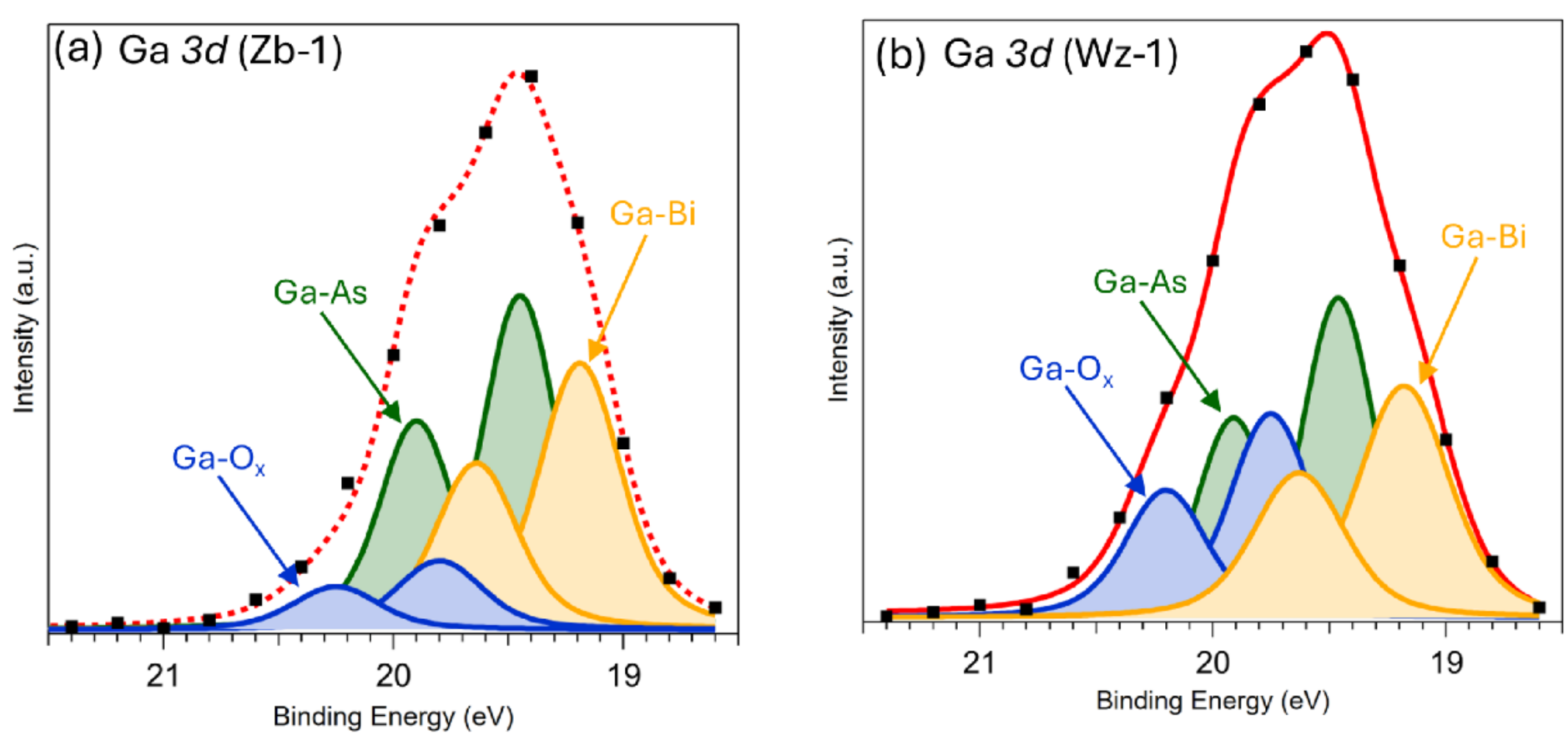

***Figure 4*** *Fitted Ga 3d core-level spectra corresponding to (a) the Zb-1 segment and (b) the Wz-1 segment shown in Fig. 2d, the intensity of the Ga-oxide component (blue peak) is smaller at the Zb and larger at the Wz segment.*

To complement our analysis of the presented nanowire XPEEM results, another NW has been analyzed in detail, as shown in Fig. S3 of the ESI, confirming the reproducibility of our results. This NW was about 100 nm thicker than the one described above. Comparable XPEEM contrast patterns and core-level spectra were observed. The only significant difference is that the thicker NW still shows a Ga-oxide component in the Ga *3d* core-level even after the extra annealing step at 300 °C. Importantly, similar trends for the Bi incorporation mechanism have been found.

## Conclusion

We have studied Bi incorporation upon radial Ga and Bi overgrowth of a GaAs heterostructure NW with axial stacking of Wz and Zb segments using synchrotron-based XPEEM. We found that Bi incorporates site-selectively by replacing As atoms via an anionic exchange reaction, resulting in a $GaAs_{1-x}Bi_x$ NW shell with larger Bi content on Zb segments and smaller Bi content on Wz segments. Bi preferentially forms Ga-Bi bonds, providing a promising approach for epitaxial III-bismide growth. Furthermore, an increased Bi content is found to protect the $GaAs_{1-x}Bi_x$ surface from enhanced oxidation upon air exposure. Our findings highlight that the crystal phases of III-V NWs govern site-selective growth, oxidation resistance, and the stability of bismide species, providing a pathway for designing quantum materials from tailored nanostructures.

## Acknowledgments:

This work was supported by the Swedish Research Council (Vetenkapsrådet, Grant No. 2021–05627), the Knut and Alice Wallenberg Foundation (KAW, Grant No. 2017.0061), the European Union through Horizon 2020 MSCA (Grant No. 945378, GenerationNano), and the NanoLund center for nanoscience (projects p18-2018 and p08-2022). We acknowledge the MAX IV Laboratory for beamtime on the MAXPEEM beamline under Proposals No. 20221449 and No. 20211193, respectively. Research conducted at MAX IV, a Swedish national user facility, is supported by Vetenskapsrådet under Contract No. 2018–07152, the Swedish Governmental Agency for Innovation Systems (Vinnova) under Contract No. 2018–04969, and Formas under Contract No. 2019–02496.

## Author declarations:

The authors declare no conflict of interest.

# Electronic Supplementary Information (ESI) for:

## Self-selective growth of $GaAs_{1-x}Bi_x$ on GaAs zinc blende/wurtzite nanowire heterostructures

Rohit Yadav[1,2,¤,*], Sebastian Lehmann[2,3], Vidar Flodgren[1,2], Evangelos Golias[4], Alexei Zakharov[4,†], Kimberly A. Dick[2,3], Anders Mikkelsen[1,2], and Rainer Timm[1,2,#]

[1]Division of Synchrotron Radiation Research, Lund University, 221 00 Lund, Sweden
[2]NanoLund, Lund University, 221 00 Lund, Sweden
[3]Centre for Analysis and Synthesis, Lund University, 221 00 Lund, Sweden
[4]MAX IV Laboratory, Lund University, 221 00 Lund, Sweden
[¤]Present Address: NNF Quantum Computing Programme, Niels Bohr Institute, University of Copenhagen, 2100 Copenhagen, Denmark

*E-mail: rohit.yadav@nbi.ku.dk
[†]deceased
[#]E-mail: rainer.timm@fysik.lu.se

### XPS fitting parameters and results:

**Table S1**: XPS peak parameters used for fitting Ga 3d core-level spectra

| | L FWHM (eV) | Spin orbital splitting (eV) | Branching ratio |
|---|---|---|---|
| Ga *3d* | 0.19 | 0.45 | 1.6 |

**Table S2**: Binding energies and peak composition (relative area of the corresponding peak component, compared to the integrated XPS signal) for Wz and Zb segments obtained from Ga *3d* peak fitting

| | | **Ga-Bi** | **Ga-As** | **$Ga-O_x$** |
|---|---|---|---|---|
| Binding Energy (eV) | Wz | 19.18 | 19.46 | 19.75 |
| | Zb | 19.19 | 19.45 | 19.80 |

### XPS thickness estimation:

The thickness of a hypothetical metallic Bi layer on top of the GaAs NW is calculated using an overlayer model [1,2] based on the XPS signal attenuation through the overlayer:

$$I_b = N_b\, \lambda_b\, \sigma_b\, e^{-\frac{d_s}{\lambda_s}}$$

$$I_s = N_s\, \lambda_s\, \sigma_s\, \left(1 - e^{-\frac{d_s}{\lambda_s}}\right)$$

$$d_s = \lambda_s \ln\left(\frac{I_s}{I_b}\frac{N_b\lambda_b\sigma_b}{N_s\lambda_s\sigma_s} + 1\right)$$

Here, $d_s$ is the calculated thickness of the overlayer and $\lambda$ is the inelastic mean free path of the photoemitted electron, based on the TPP-2M formula. [3] The $I_b$ and $I_s$ are the intensity of the XPS signal from the bulk and the overlayer, respectively, $\sigma$ is the ionization cross-section, and $N$ is the number of atoms per unit volume, calculated as $N = A_v * \rho/M$. Here $A_v$ is Avogadro's number, $\rho$ is the density, and $M$ is the molecular weight of the material.
The values used for calculation are shown in Table S3.

**Table S3:** Values used for evaluating the Bi thickness at 80 eV photon energy based on the overlayer model

| | GaAs (bulk) | Bi (surface) |
|---|---|---|
| $E_g(eV)$ | 1.42 | 0 |
| $\lambda$ (Å) | 4.77 | 5.65 |
| $N(\text{Å}^{-3})\ x\ 10^3$ | 22 | 28.2 |
| $\sigma\ (Mbarn)$ [4] | 8.39 | 11.07 |

## Additional PEEM maps and intensity profiles:

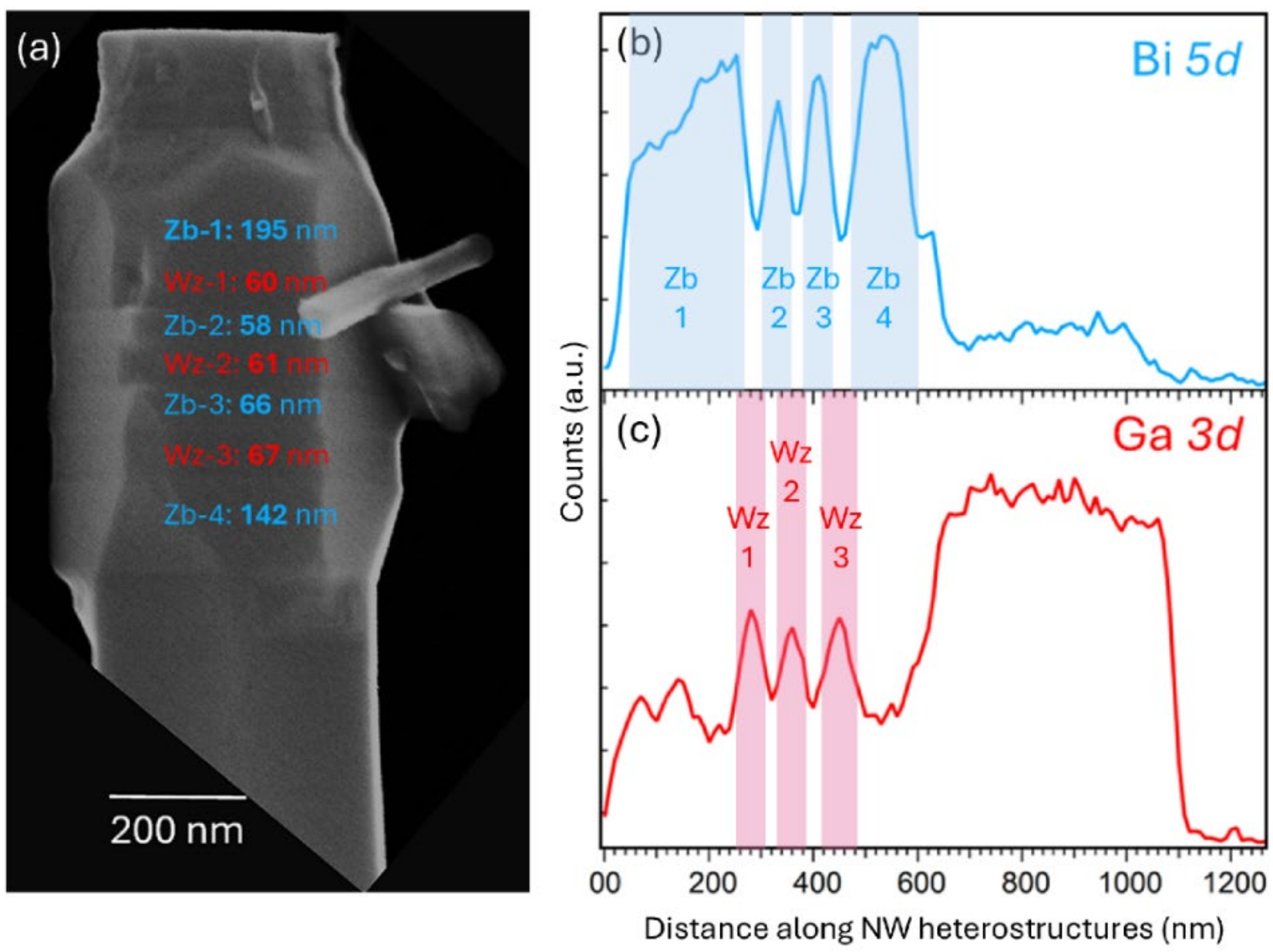

*Figure S1 (a) SEM image of the analyzed NW, overlaid labels indicate the width of the corresponding Wz or Zb segment extracted using the image contrast. (b,c) Intensity line profiles from the (b) Bi 5d and (c) Ga 3d PEEM maps shown in Fig.2 of the main manuscript, extracted from the NW top to bottom, i.e., across the heterostructure. The vertical stripes indicate the segments with high Bi signal intensity (blue) and Ga signal intensity (light red) in the PEEM maps, which correlate with the positions of Zb and Wz segments, according to (a).*

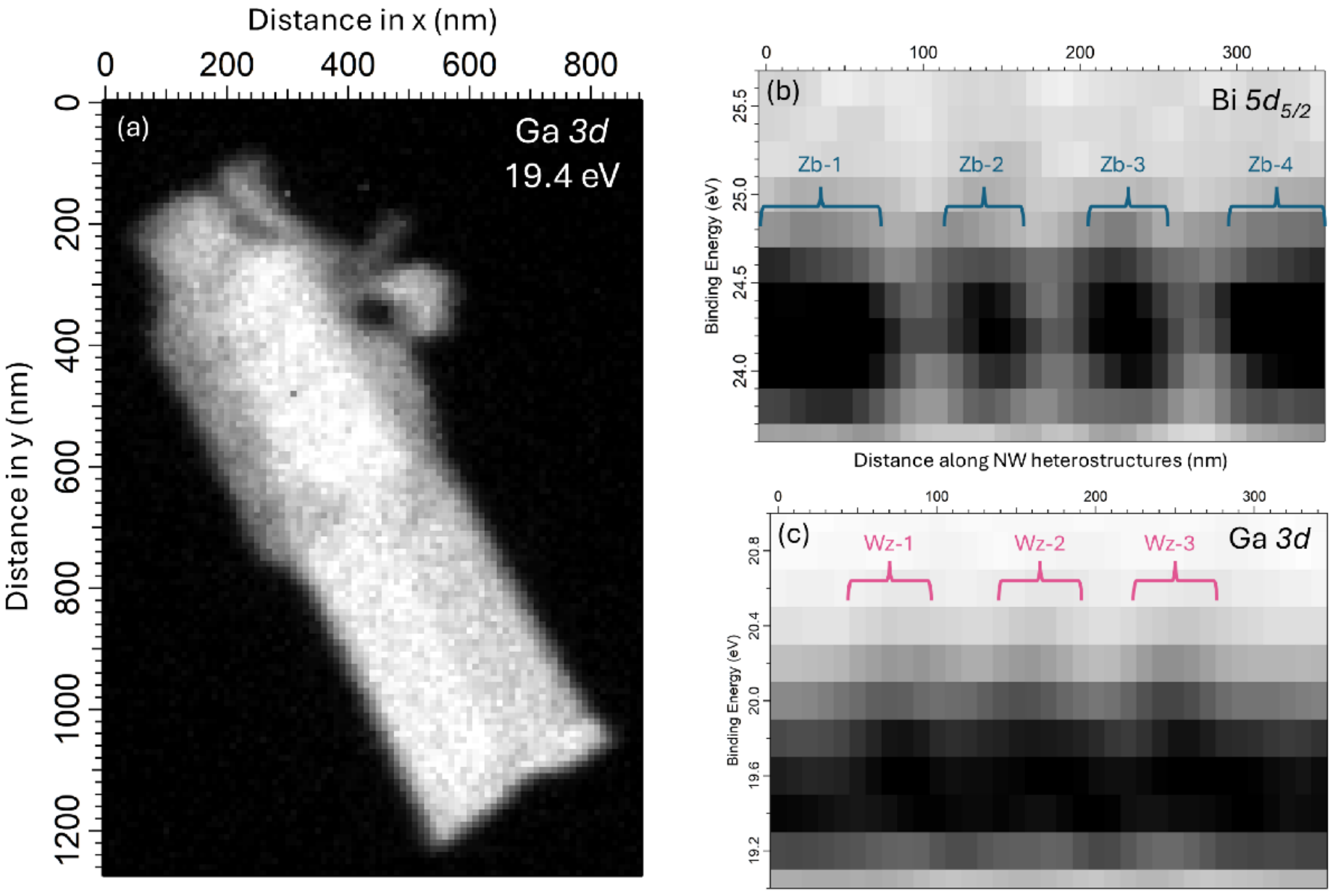

*Figure S2 (a) Ga 3d PEEM map of the same NW as shown in Fig. 2 of the main manuscript, here obtained at 19.4 eV BE, where the contrast between Wz and Zb segments of the heterostructure is indistinguishable due to their similar XPS intensity. (b,c) Maps of BE vs distance along the NW heterostructure extracted from a series of PEEM maps at various BEs corresponding to (b) the Bi $5d_{5/2}$ core level and (c) the Ga 3 core level; the different segments of the NW heterostructure are indicated.*

## PEEM analysis of another NW:

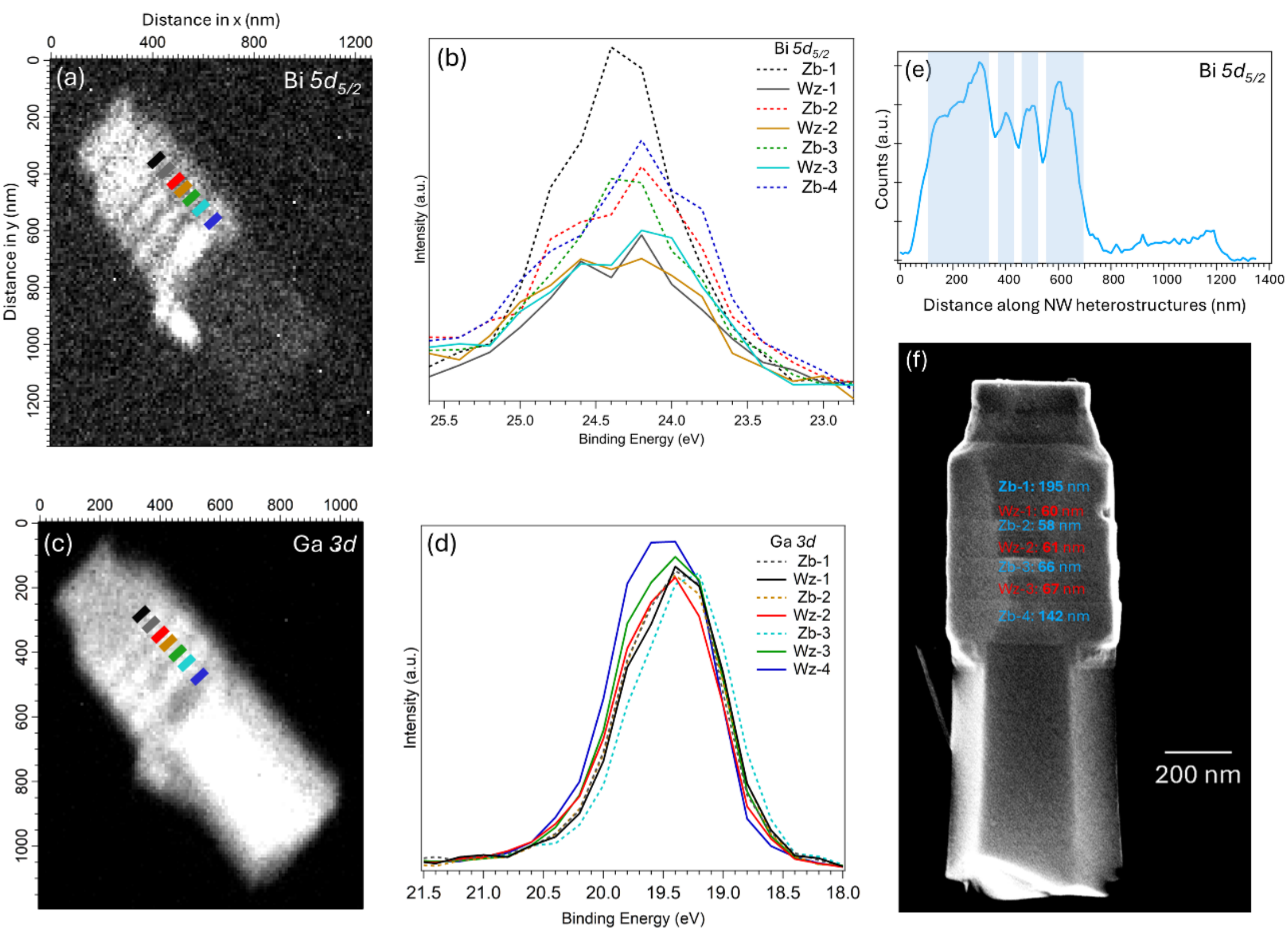

*Figure S3 PEEM (a) Bi 5d and (c) Ga 3d core-level maps of the same NW and corresponding (b) Bi $5d_{5/2}$ and (d) As 3d core-level spectra, extracted from series of PEEM maps with slightly varying BE, for different color-coded regions in (a,c). The photon energy used for PEEM mapping is 80 eV. (e) Intensity line profile from the Bi 5d PEEM map in (a), extracted from the NW top to bottom; blue vertical stripes indicate the segments with high Bi signal intensity, which correlate with the positions of Zb segments. (f) SEM image of the same NW, the overlaid labels indicate the width of the corresponding Wz and Zb segments.*